\documentclass[twocolumn]{aastex631}
\usepackage{comment}
\usepackage{gensymb}
\usepackage{textcomp}
\usepackage{float}
\usepackage{amssymb}
\usepackage{appendix}
\usepackage[normalem]{ulem}
\usepackage{cancel}
\usepackage{xcolor}
\usepackage{soul}
\usepackage{multirow}
\usepackage{amsmath}
\usepackage{graphicx}
\usepackage{cleveref}

\shortauthors{Saha et al.}
\graphicspath{{./figures/}}

\definecolor{forestgreen}{RGB}{34,139,34}

\definecolor{mypink}{cmyk}{0, 0.7808, 0.4429, 0.1412}

\begin{document}
%\title{Generating Kilonova Light Curves Using Autoencoder to Investigate the Properties of the Binary Merging System  }
\title{Rapid Generation of Kilonova Light Curves Using Conditional Variational Autoencoder }

\correspondingauthor{Surojit Saha}
%\correspondingauthor{Albert K.~H. Kong}
 
\email{surojitsaha@gapp.nthu.edu.tw}
%\email{akong@gapp.nthu.edu.tw}
\author[0000-0002-3333-8070]{Surojit Saha}
\affiliation{Institute of Astronomy,\\ National Tsing Hua University, Hsinchu City, Taiwan (R.O.C.) }
\author[0000-0003-2198-2974]{Michael J. Williams}
\affiliation{Institute for Gravitational Research, \\School of Physics and Astronomy, University of Glasgow, \\Glasgow, Scotland, UK}
\affiliation{Institute of Cosmology and Gravitation, University of Portsmouth, Portsmouth PO1 3FX, UK}
\author[0000-0002-0290-3129]{Laurence Datrier}
\affiliation{Institute for Gravitational Research, \\School of Physics and Astronomy, University of Glasgow, \\Glasgow, Scotland, UK}
\author[0000-0001-7628-3826]{Fergus Hayes}
\affiliation{Institute for Gravitational Research, \\School of Physics and Astronomy, University of Glasgow, \\Glasgow, Scotland, UK}
\author[0000-0002-2555-3192]{Matt Nicholl}
\affiliation{Astrophysics Research Centre, School of Mathematics and Physics, Queens University Belfast, Belfast BT7 1NN, UK}

\author[0000-0002-5105-344X]{Albert K.~H. Kong}
\affiliation{Institute of Astronomy,\\ National Tsing Hua University, Hsinchu City, Taiwan (R.O.C.) }
\author[0000-0001-8322-5405]{Martin Hendry}
\affiliation{Institute for Gravitational Research, \\School of Physics and Astronomy, University of Glasgow, \\Glasgow, Scotland, UK}
\author[0000-0002-1977-0019]{IK Siong Heng}
\affiliation{Institute for Gravitational Research, \\School of Physics and Astronomy, University of Glasgow, \\Glasgow, Scotland, UK}
\author[0000-0001-5169-4143]{Gavin P. Lamb}
\affiliation{Astrophysics Research Institute, Liverpool John Moores University, IC2 Liverpool Science Park, 146 Brownlow Hill, Liverpool L3 5RF}
\author[0000-0002-0030-8051]{En-Tzu Lin}
\affiliation{Institute of Astronomy,\\ National Tsing Hua University, Hsinchu City, Taiwan (R.O.C.) }
\author[0000-0003-3772-198X]{Daniel Williams}
\affiliation{Institute for Gravitational Research, \\School of Physics and Astronomy, University of Glasgow, \\Glasgow, Scotland, UK}

\begin{abstract}
The discovery of the optical counterpart, along with the gravitational waves from GW170817, of the first binary neutron star merger, opened up a new era for multi-messenger astrophysics. Combining the GW data with the optical counterpart, also known as AT2017gfo, classified as a kilonova, has revealed the nature of compact binary merging systems by extracting enriched information about the total binary mass, the mass ratio, the system geometry, and the equation of state. Even though the detection of kilonova brought about a revolution in the domain of multi-messenger astronomy, since there has been only one kilonova from a gravitational wave detected binary neutron star merger event so far, this limits the exact understanding of the origin and propagation of the kilonova. Here, we use a conditional variational autoencoder trained on light curve data from two kilonova models  having different temporal lengths, and consequently, generate kilonova light curves rapidly based on physical parameters of our choice with good accuracy. Once trained, the time scale for light curve generation is of the order of a few milliseconds, thus speeding up generating light curves by $1000$ times compared to the simulation. The mean squared error between the generated and original light curves is typically $0.015$ with a maximum of $0.08$ for each set of considered physical parameter; while having a maximum of $\approx0.6$ error across the whole parameter space. Hence, implementing this technique provides fast and reliably accurate results.
\end{abstract}

\section{Introduction} \label{sec:intro}
The concomitant discovery of the gravitational waves (GW) and optical counterpart electromagnetic waves (EM) \citep{2017ApJ...848L..16S, Lipunov_2017,Tanvir_2017,Arcavi_2017,2017ApJ...848L..24V, doi:10.1126/science.aap9811} from the merger of the binary neutron star GW170817 \citep{PhysRevLett.119.161101}, also accompanied by short gamma-ray burst (GRB) \citep{2017ApJ...848L..14G,Savchenko_2017}, advanced the domain of multi-messenger astronomy. This optical counterpart, designated as kilonova (KNe) and known as AT2017gfo bolstered previous predictions of the existence of such electromagnetic transients \citep{Li_1998,1999A&A...341..499R}. Prior to the discovery of GW170817, it was predicted that such an event would be accompanied by an EM counterpart, including short-duration GRBs
\citep{1989Natur.340..126E,2007PhR...442..166N}, emissions ranging from radio to X-rays pertaining to on or off axis afterglow \citep{van_Eerten_2012,2014MNRAS.445.3575C,2015ApJ...815..102F,Lamb_2016} and optical to near-IR resulting from the decay of r-process nuclei.
A kilonova is an isotropic quasi-thermal transient which is powered by the radioactive decay, in the merger ejecta, of r-process nuclei, having luminosities $10^{40}-10^{42}\,$erg\,s$^{-1}$
\citep{Li_1998,1999A&A...341..499R,2010MNRAS.406.2650M,2013ApJ...775...18B}. This EM counterpart can provide deeper understanding of the merger environment and its products. When EM information is further combined with GW, it leads to a unique platform for an extensive  understanding of such binary events.  
KNe models consists of one or more radioactive ejecta components that produce light curves peaking at different timescales and temperature depending on  atomic mass number of the ejecta, and luminosity. The two-components consists of the blue KNe \citep{2010MNRAS.406.2650M,2011ApJ...736L..21R,2014MNRAS.441.3444M} emission having poor lanthanide fraction $(10^{-5})$ in the merger ejecta peaking at a relatively earlier time scale and red KNe \citep{2013ApJ...775...18B,Kasen_2013,Tanaka_2013} comprising of lanthanide-rich $(10^{-2})$ merger ejecta with peak values at later days.
In the three-component model \citep{2017} there is an inclusion of purple KNe that describes the presence of purple KNe with a lanthanide fraction of $10^{-3}$. Various papers \citep{2017ApJ...848L..17C,2018ApJ...862L..11V} have provided the best-fit parameter for AT2017gfo related to the blue, purple and red kilonova in terms of ejecta mass, ejecta velocity and lanthanide fraction. 
However, since there has been only one GW-confirmed detection of a KNe from a merger of binary neutron stars, along with some possible candidates from other sources, it is difficult to understand and verify the properties of binary merging systems that emit electromagnetic radiation. An overview  on the physical parameters that govern the KNe has been provided by \citet{Metzger_2019}.

In recent years, machine learning has been used for various data analysis and formulating techniques required 
in astronomy \citep{2014AAS...22325301V,2010IJMPD..19.1049B,2021arXiv211114566N,https://doi.org/10.48550/arxiv.2202.07121}. There has been certain areas where the application of machine learning techniques has produced a remarkable result, especially focusing on providing a relatively faster results alternative to the existing
methods~\citep{2011arXiv1110.5688B,2020AAS...23510915S}.
There are plenty of excellent literature available on different topics of machine learning application in astronomy and astrophysics spanning over wide range of topics
\citep{2022MNRAS.509.2289L,2022MNRAS.509..990G,2022arXiv220506758G,2022MNRAS.512.5580S}. 
In this paper, we incorporate a method from ML domain known as autoencoder \citep{1986Natur.323..533R}, which is based on feed-forward mechanism, to generate light curves. The primary task of an autoencoder is to encode the input into a lower dimensional latent representation and then decode it back to data. In feed-forward mechanism, the neural network has unidirectional processing of information. The striking feature for an autoencoder is that the encoder compresses the high-dimensional  input to a low-dimensional latent space and the decoder decompressed to produce result to match the input. 

In this work, we regenerate the KNe light curves by implementing a conditional variational autoencoder (CVAE) \citep{Kingma_2019}, which is developed based on variational autoencoeder (VAE)\citep{2013arXiv1312.6114K}. We use this CVAE to generate light curves based on our choice of physical parameter values of the binary merging system. For the CVAE, once the training has been completed, we are more flexible to rapidly generate light curves based on the physical parameters depending on choice of the data set.  Although, KNe data from only two models for training and producing the results are used in this work, it is also possible to extend the same idea for data from other models. Although the data have been produced from the model, this unique data analysis technique can not only generate light curves for the parameter values that are not explicitly mentioned in the model but can also replace time-consuming and resource-draining simulations required for predicting the light curves. The novelty of this technique is expressed in the fact that once the CVAE is trained on a KNe model, we can generate numerous light curves for a different combinations of physical parameters in a very short time. 
 An alternative approach for such conditional variational autoencoder and KNe work has been presented in \citet{https://doi.org/10.48550/arxiv.2204.00285} where different method has been studied for obtaining results. One of the fundamental differences between \citet{https://doi.org/10.48550/arxiv.2204.00285} and our work is in using spectra rather than light curves during the training. In our method, we look into the light curve evolution with respect to different sets of physical parameters aiming to complete the process in a relatively shorter time-scale compared to the existing simulation methods. This kind of rapid generation technique for KNe light curves can be useful when used as template for rapid parameter estimation of KNe.
 
In the following sections, we show the gradual implementation of our idea, and the results are discussed thereafter. In Section~\ref{sec:autoencoder}, we discuss the CVAE architecture implemented in our work. Section~\ref{sec:data} puts forward the differences in the data and their physical parameters used for training and generation of the KNe light curves. Section~\ref{sec:result} provides the detailed discussion of results obtained after implementation of CVAE. In Section~\ref{sec:dis} we summarize our approach and present some important features of this technique. Additional results for references are included in the appendix. 
\section{Kilonova Models}
\label{sec:knmodels}
Kilonova emission results from the mass ejection in the NS mergers \citep{1999ApJ...525L.121F,refId0,PhysRevD.87.024001}. The properties of the ejecta such as the ejecta mass, ejecta velocity and opacity dominates the peak luminosity and the time of the peak luminosity \citep{Li_1998,Kasen_2013,10.1093/mnras/stv721,2013ApJ...775...18B,2017PASJ...69..102T}. This luminosity is sourced from the radioactive decay of the r-process elements synthesized in the merger. 
This work primarily focuses on two KNe models having different physical parameters that determine the light curves. Below we provide an outline of the two models used in this work. In \citet{2017Natur.551...80K},the light curves are dependent upon the peak magnitude and decay time with respect to the ejecta mass, velocity and lanthanide fraction is provided. 
Quantitatively, peak luminosity to first order is denoted by $L\propto M_{\rm ej}$ and the width of the light curve is $\tau \propto (\kappa M_{\rm ej}/v_{\rm ej})^{1/2}$, pointing to  lanthanide-rich ejecta with higher $\kappa$, where $M_{\rm ej}$, $v_{\rm ej}$ and $\kappa$ are the mass, velocity and opacity of the ejecta. Hence heavier ejecta from the merger leads to higher peak magnitude with relatively longer duration KNe whereas ejecta with higher velocities have short duration with bright KNe. We see that lanthanide fraction plays a major role on the light curves where the ejecta with lower lanthanide fraction decays on a relatively shorter timescale compared to lanthanide-rich ejecta decaying over weeks. From the literature in \citet{2021MNRAS.505.3016N}, the study presents an approach where light curves are predicted from the chirp mass, mass ratio, orbital inclination also including the properties related to the nuclear equation of state. Here we see the nature of light curves based on these physical parameters. It is important to note that, \citep{2017Natur.551...80K} is based on radiative transfer model, where as \citep{2021MNRAS.505.3016N} is a semi-analytic model. While KNe light curves can be generated with CVAE, we choose these models due to their different parametrizations and the data availability.  However, this does not put any limitations on the use of CVAE, since data from any other available KNe models can be equally used.

\section{Autoencoder}
\label{sec:autoencoder}
\begin{figure*}[!ht]
\plotone{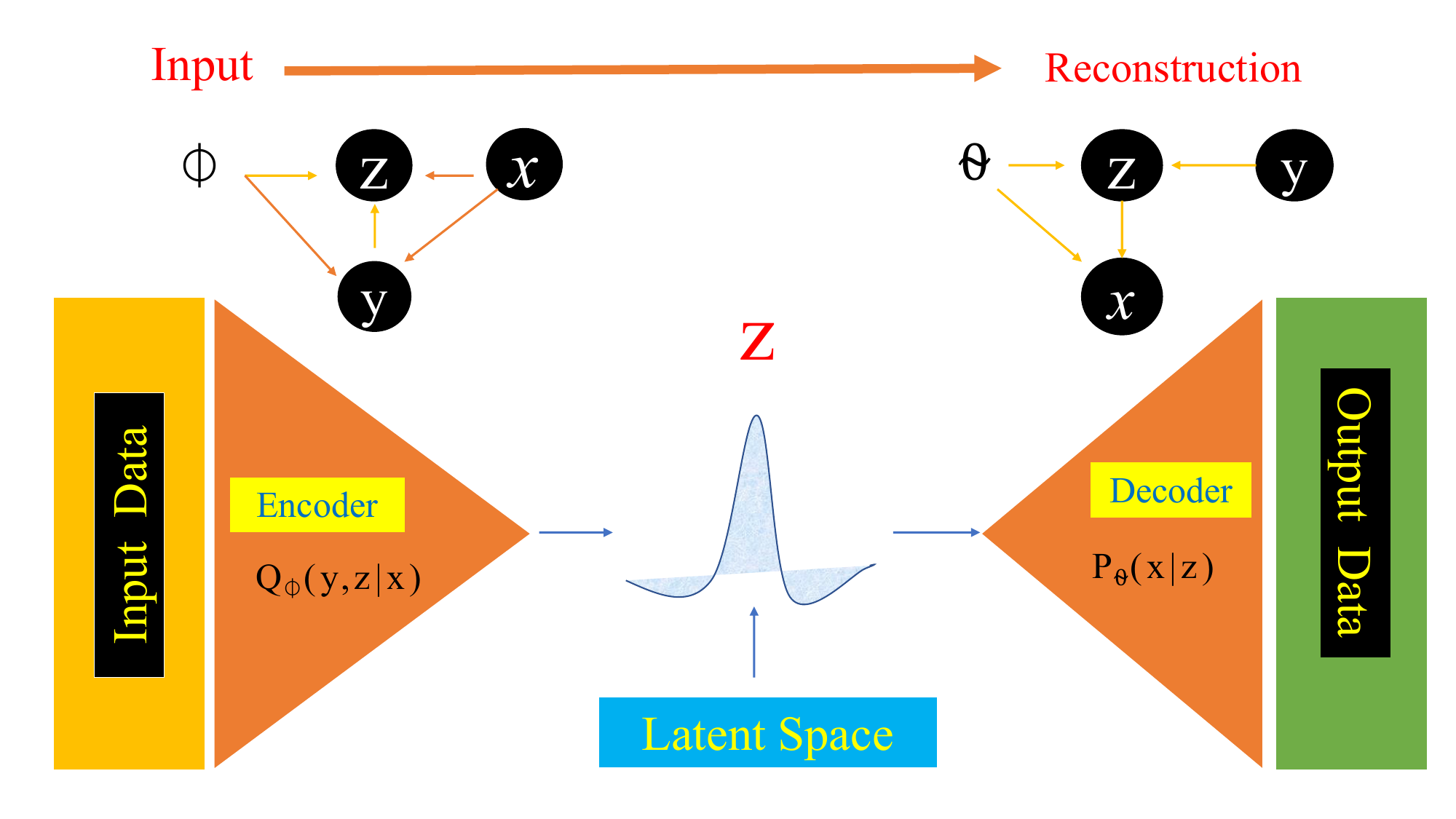}
\epsscale{1.2}
\caption{Schematic diagram of the conditional variational autoencoder, consisting of the encoder, latent space and the decoder, from input to data generation is shown. In the figure, we have the probabilistic encoder $Q(\phi)$ and the probabilistic decoder $P(\theta)$. \emph{Z} represents the compressed input or encoded representation. Here, the encoder transforms the  high-dimensional input data, into the low-dimensional latent space which represents the compressed form of the high-dimensional input data while the decoder maps this compressed representation in the low-dimensional latent space leading to the reconstruction of the original high-dimensional data. We take a set of data \emph{x}, which is the KNe light curves and condition it on a given parameter \emph{y}, which in our case are the physical parameters from the model, and then the encoder compresses this representation in the latent space. The decoder decompresses the representation and provides the relevant output data, KNe light curves, based on the physical parameter values of our choice. This is an overview of the CVAE architecture that has been used in this work to train and generate the KNe light curves.} 
\label{fig:cvae}
\end{figure*}

 With the advancement of new machine learning (ML) methods and their implicit application in data analysis, has opened a new era where ML techniques can be implemented to obtain relatively faster results. Time-consuming and resource-draining simulations can be completed on a reasonable time scale \citep{RevModPhys.91.045002}. There are many available ML techniques that can be used for data analysis with specific techniques built to obtain certain results. We use an autoencoder, an ML technique that is based on a feed-forward mechanism, and its function is to reproduce the input. 
As is the case with any standard autoencoder, the CVAE consists of three sections, an encoder, a latent space, and a decoder, as shown in Fig.~\ref{fig:cvae}. The encoder $(Q(\phi))$ compresses the high-dimensional input data into a lower-dimensional latent space while capturing the features of the input data. Latent space(\emph{Z}), which is also referred to as the output of encoder, has a lower dimensional representation. Generally, a well-trained CVAE has the entire high-dimensional input data smoothly distributed over the latent space. To generate light curves of our choice, we draw samples from this latent space and pass it through the decoder where the decoder $(P(\theta))$ takes the compressed representation and generates the required light curves. This facilitates using CVAE as a tool, not only limited to a generative model but also quickly look into the parameter dependency of KNe light curves. In CVAE, the training is regularised in order to avoid over-fitting. Hence, the latent space is well distributed to enable generative process 
Even though there are other generative models like Generative Adversarial Networks (GAN)\citep{goodfellow2014generative} but since in our work we want to sample new data from the probability distribution of the input data, to look at the model parameters crucial for the light curves, we prefer to utilize the strength of an autoencoder because of its ability to have a probability distribution of input.
\par
Even though GAN and CVAE are both generative models and belongs to the category of unsupervised learning, one of the primary differences lies in there architecture where GAN has a generator and discriminator and CVAE has encoder, latent space and decoder. Along with this, the loss functions used in GAN and CVAE are different from each other. It is important to note that for generating new data, sampling a hidden state in GAN takes place from a predefined distributions and is then fed into the discriminator in contrast to that of CVAE where sampling of hidden state, to be fed into the decoder, is done from selecting a prior distribution related to the actual data. 
Here, to achieve our goal, we take advantage of the conditional variational autoencoder, where we train it on the light curves by conditioning them on physical parameters and generate new light curves based on the physical parameters of our choice. In this case, we have complete control over the generated data set. In this work, we take advantage of this feature of CVAE to generate new data.  In our CVAE architecture, the loss function is the combination of the reconstruction-loss and the Kullback-Leibler divergence  \citep{2020arXiv200207514A}.
Training is carried out with $batch\_size=50$ for $epochs=1000$ and having a learning rate of $0.0001$. Commencing from the training till generation of the light curves for the different physical parameter values, it takes $\simeq$ 20 minutes for the process to complete using $3.9$ GHz $8$-core Intel Core i$9$ processor with $32$ GB memory. In our training and regeneration of the light curves, we have not used the GPU of the system and have completely relied on the CPU. Once the model is trained and saved, it takes only a few milliseconds to generate the light curves. In this section, we outline the CVAE without delving into the details, however, for more detailed variational encoder readers can refer to \citep{2013arXiv1312.6114K} and \citep{Kingma_2019}. In  Section~\ref{sec:result} more insight into the results obtained from CVAE are shown.
\par

\begin{figure*}[!ht]
\gridline{
          \fig{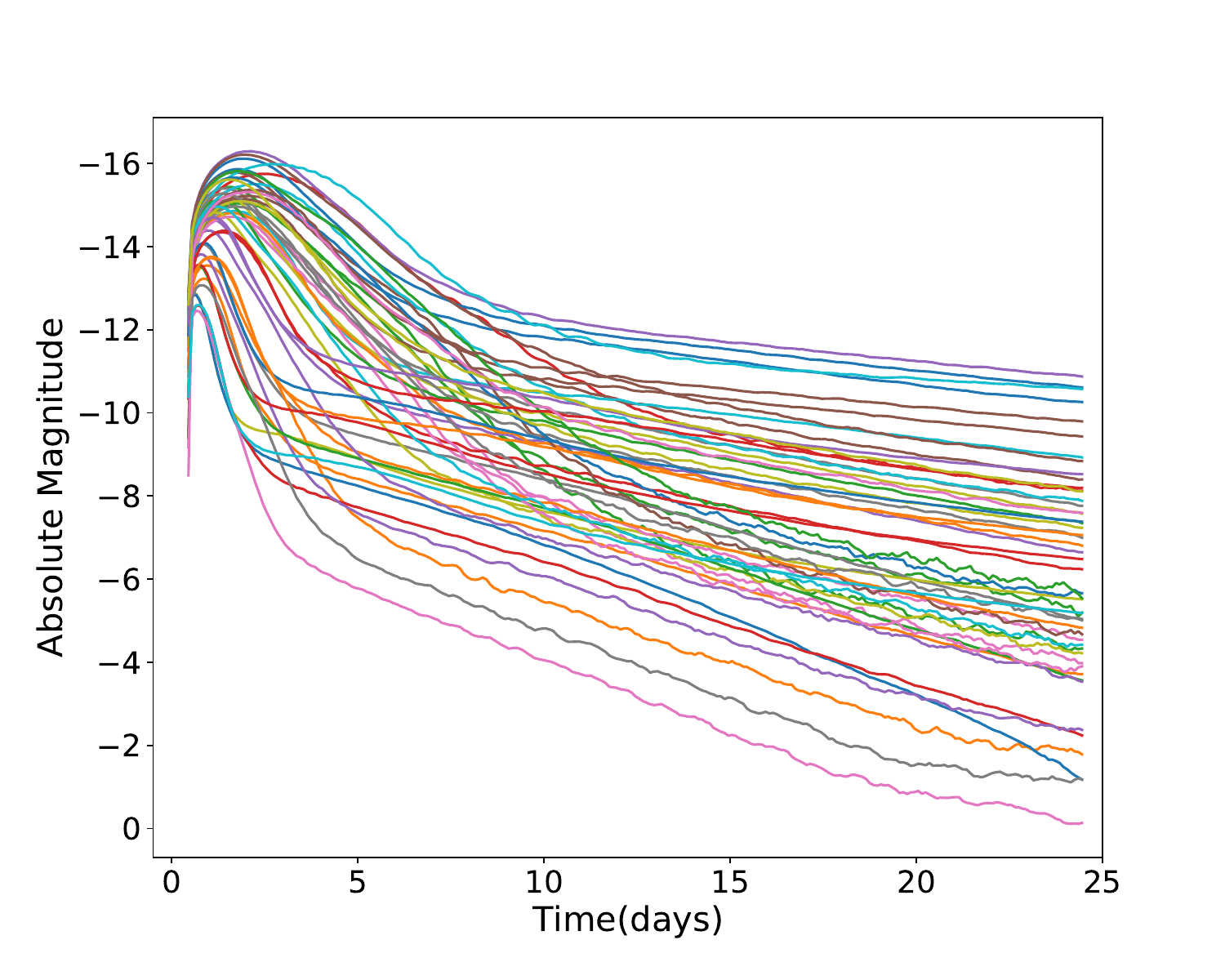}{0.5\textwidth}{(a)\hspace{0.1cm}Input for $D_{1}$}\label{fig:input_scale}
          \fig{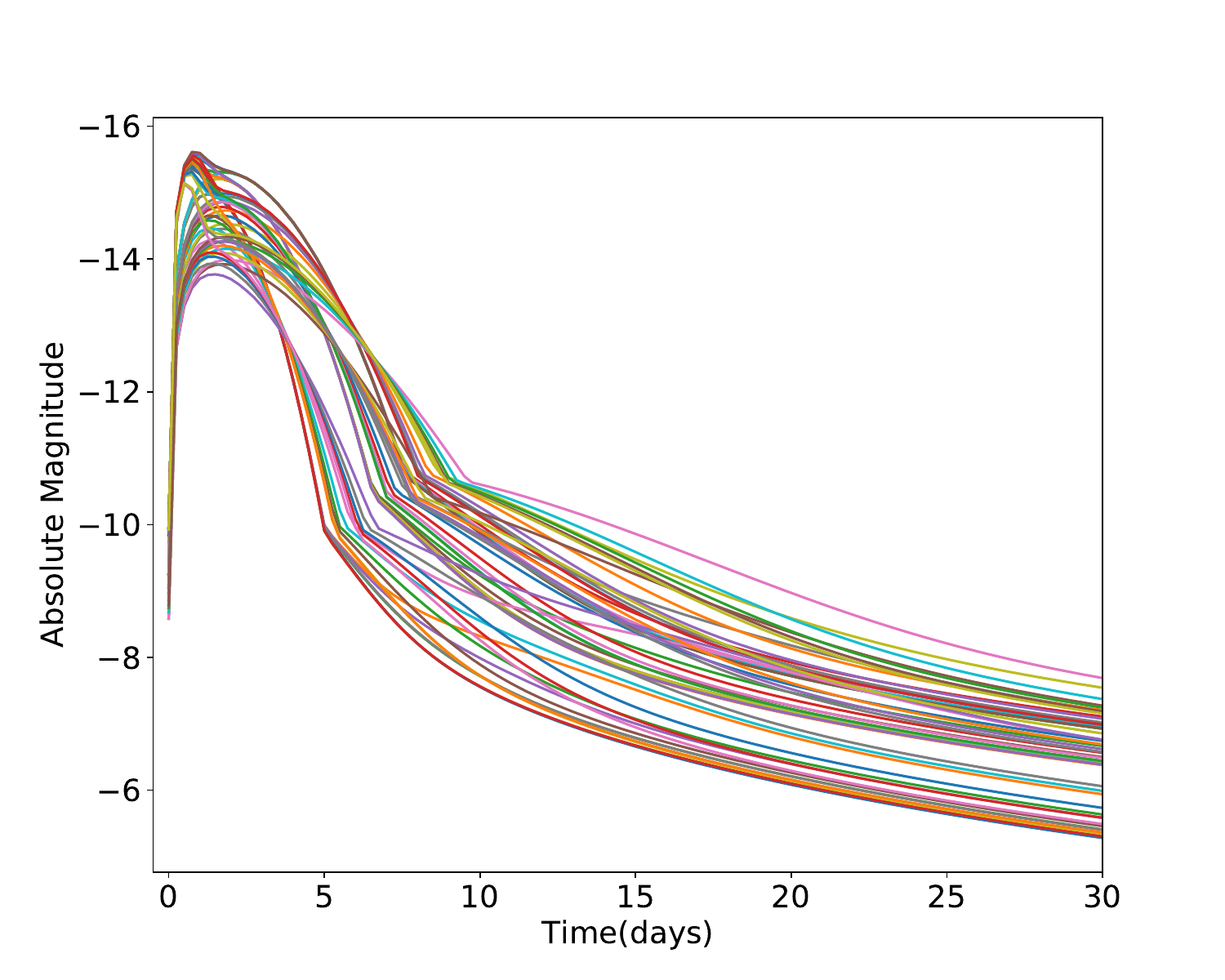}{0.52\textwidth}{(b)\hspace{0.1cm}Input for $D_{2}$}}
\caption{(a) In this plot, we represent only $50$ KNe light curves of the g-band of $D_{1}$ which are a part of the data set for training. These light curves consist of the mass of the ejecta, the velocity of the ejecta, and the fraction of lanthanide as physical parameters with values ranging between ($0.001-0.1M_{\odot}$),$(0.03-0.05c)$ , $(10^{-9}-10^{-4})$ respectively. (b)In this plot, we show $50$ light curves for g-band of $D_{2}$ which are a part of the training data set. These light curves consists of chirp mass, mass ratio, fraction of the remnant disk with values ranging between ($0.7-1.10M_{\odot}$), $(0.50-1.0)$, $(0.15-0.45)$ respectively and having a viewing angle of $0\degree$ as their physical parameters. This figure gives an outline of the KNe light curves and their decay time, which is used for training the CVAE. For the both data, the KNe light curves are in absolute magnitude. These light curves are scaled between [$0$-$1$] while feeding them into the CVAE for training, but, while representing the generated results, these light curves are scaled back to the absolute magnitude.}
\label{fig:inputall}
\end{figure*}
    
\section{Data}\label{sec:data}    

In the ML domain, one needs to be very careful about the data that is fed into the algorithm. The data have to be pre-processed and prepared accordingly. The data used for training, validation and  test set are categorized into two types as mentioned in Section~\ref{sec:knmodels}. We adopt the data (\url{https://github.com/dnkasen/Kasen_Kilonova_Models_2017}), hereafter $D_{1}$ \citep{2017Natur.551...80K} to prepare our data set to be fed into the CVAE. The physical parameters in this data set are ejecta mass ($0.001-0.1M_{\odot}$), ejecta velocity($0.1-0.3c$) and lanthanide fraction(10\textsuperscript{-4}-10\textsuperscript{-9}). 
There are $329$ light curves with duration of $\approx25$ days. Each light curve has different values of the physical parameter within the range mentioned above. 
The light curves are available for different filter bands (u,g,r,y,z) within the same physical parameter range. Using these data, $241$ light curves are used as training sets, and the rest are equally divided into the test set and the validation set. In our analysis, we have trained the CVAE on each filter band data separately, hence there are five trained CVAE's, and carried out light curves generation; however, in the main sections, only the g-band results are shown; the rest are added in appendix.
\par
The second type of data, hereafter $D_{2}$, used here consists of simulated light curves at same the filter bands but having different  physical parameters. The physical parameters of these data are the chirp mass of the binary system, the mass ratio, the fraction of the remnant disk that is ejected, the viewing angle in degrees from the pole, and the opening angle of the cocoon shock \citep{2021MNRAS.505.3016N}. The range of values are $(0.7M_{\odot}-2.0M_{\odot})$ for the chirp mass, $(0.5-1.0)$ for the mass ratio, $(0.15-0.45)$ for the fraction of the remnant disk, respectively. For viewing angle the data corresponds to the values of $0\degree$, $60\degree$ and $90\degree$. There are $529$ light curves having temporal length of $30$ days, out of which $401$ light curves are included in the training data and the rest are equally divided into test and validation set.
The difference in the two data set is appears in the physical parameters used in the respective KNe models. These data are used to verify the pliability of the CVAE. Since the physical parameters in the two data sets used are entirely different, this gives an extra edge in verifying our approach, and the results are detailed in Section~\ref{sec:result}.
Since both data sets contain light curves that have different absolute magnitude peak values, while feeding them for training, we have scaled the light curve data and the physical parameters between $0$ and $1$, as this would be a more effective approach.
Later, after training and generating light curves, we rescale the light curves and plot accordingly. 
Throughout the paper, we will follow a particular format to represent the physical parameters in the text and in the legends of the plots. For the texts and plots relevant to $D_{1}$, we will use the format 
where $a$,$b$ and $c$ will be the values of ejecta mass in $M_{\odot}$,
velocity of the ejecta in units of velocity of the light and lanthanide fraction respectively. Similarly for the $D_{2}$, we will use the format of $[w,x,y,z]$ where $w$ is the chirp mass in units of $M_{\odot}$, $x$ is the mass ratio, $y$ gives the fraction of the remnant disk that is ejected and $z$ is the viewing angle in degrees from the pole. We will use the above format throughout the paper wherever required, without further mentioning the respective units. Fig.~\ref{fig:inputall}(a) shows $50$ out of $329$ original light curves from $D_{1}$ in the g-band for physical parameters corresponding to ejecta mass, ejecta velocity and lanthanide fraction with values ranging between $0.001-0.1M_{\odot}$, $0.03-0.05c$ and $10^{-4}-10^{-9}$ respectively whereas in Fig~\ref{fig:inputall}(b), we show $50$ out of $529$ light curves corresponding to $D_{2}$ having physical parameters that of chirp mass ($0.7-1.10M_{\odot}$), mass ratio ($0.5-1.0$), fraction of the remnant disk ($0.15-0.45$) and having a viewing angle of $0\degree\hspace{0.05cm}$ from the pole as mentioned in Section~\ref{sec:data}. In Fig.~\ref{fig:inputall}, we show the input light curves corresponding to $D_{1}$ and $D_{2}$ which is fed into the CVAE. During training and validation, scaled values of the KNe light curves and the conditional physical parameters are used for both $D_{1}$ and $D_{2}$, and the obtained results are shown in absolute magnitude values. The input light curves for training from $D_{1}$ consist of $241$ light curves and from $D_{2}$ has $401$ light curves from the g-band. As previously mentioned, each light curve has different physical parameters corresponding to ejecta mass, ejecta velocity and lanthanide fraction for $D_{1}$ and chirp mass, mass ratio, fraction of the remnant disk and the viewing angle from the pole for $D_{2}$. Throughout the text, $D_{1}$ and $D_{2}$ will be referred to accordingly, keeping the above training, test, and validation set unchanged. In Table~\ref{table:1}, we tabulate the physical parameters and the data ranges used in this work. In this study, while defining the training, test, and validation set we do not restrict the physical parameters to the different regions of the parameter space. Instead, the physical parameters present in the training, test, and validation sets cover the entire parameter space. But, at the same time, we ensure that none of the parameter combinations, i.e ejecta mass, ejecta velocity, and lanthanide fraction for $D_1$ and chirp mass, mass ratio, a fraction of the remnant disk, and viewing angle for $D_2$, are repeated in the training, test, and validation set. Specifically, for $D_2$, apart from the training, test, and validation set, we have kept aside a separate set of simulated light curves from MOSFiT(\url{https://github.com/guillochon/MOSFiT}), which are utilized to evaluate the performance of the CVAE. We augmented the existing test set with these separately simulated light curves, hereafter $D_2\dagger$, to provide a robust performance check. The combination of the physical parameters present in $D_2\dagger$, as shown in Table 2, covers the parameter space but is previously unseen to the CVAE. The CVAE will be employed with these sets of physical parameters to generate light curves and hence compare with the true light curves, thus providing a robust approach to appraise the CVAE performance.
Hence, using the CVAE, we generate light curves over a grid on the parameter space.

\begin{table*}[ht!]
  \centering
  \renewcommand{\arraystretch}{3}
  \resizebox{6in}{!}{%
    \begin{tabular}{|c|c|c|}
      
      \hline
      \multicolumn{1}{|c|}{\textbf{\LARGE Data}} & \multicolumn{1}{c|}{\textbf{\LARGE Physical Parameters}} & \textbf{\LARGE Data Features} \\
      
      \hline
      \multicolumn{1}{|c|}{\multirow{3}{*}{\textbf{\Huge $D_1$}}} & \textbf{\LARGE Ejecta Mass ($M_{\odot}$)} & \LARGE $0.001$, $0.0025$, $0.005$, $0.01$, $0.02$, $0.025$, $0.03$, $0.04$, $0.05$, $0.075$ \\
      \cline{2-3}
      & \textbf{\LARGE Ejecta Velocity(c)} & \LARGE $0.03$, $0.05$, $0.1$, $0.2$, $0.3$ \\
      \cline{2-3}
      & \textbf{\LARGE Lanthanide Fraction} & \LARGE $10^{-9}$, $10^{-5}$, $10^{-4}$, $10^{-3}$, $10^{-2}$, $10^{-1}$     \\
      \hline
      \hline
      \multicolumn{1}{|c|}{\multirow{4}{*}{\textbf{\Huge $D_2$}}} & \textbf{\LARGE Chirp Mass ($M_{\odot}$)} & \LARGE $0.7$, $0.8$, $0.9$, $1.0$, $1.1$, $1.2$, $1.3$, $1.4$, $1.5$, $1.6$, $1.7$, $1.8$ \\
      \cline{2-3}
      \multicolumn{1}{|c|}{} & \textbf{\LARGE Mass Ratio} & \LARGE $0.5$, $0.6$, $0.7$, $0.8$, $0.9$, $1.0$ \\
      \cline{2-3}
      \multicolumn{1}{|c|}{} & \textbf{\LARGE Fraction of the Remnant Disk} & \LARGE $0.15$, $0.30$, $0.45$ \\
      \cline{2-3}
      \multicolumn{1}{|c|}{} & \textbf{\LARGE Viewing Angle} & \LARGE $0\degree$, $60\degree$, $90\degree$ \\
      \hline
    \end{tabular}%
  }
\caption{{This table gives an overview of the physical parameters and range of their values corresponding to $D_{1}$ and $D_{2}$ used in this work for training and generation of the KNe light curves. $D_{1}$ is taken from \url{https://github.com/dnkasen/Kasen_Kilonova_Models_2017} and $D_{2}$ is adopted from \url{https://github.com/mnicholl/kn-models-nicholl2021}. The training, test and validation data for the corresponding data set consists of these individual physical parameter.
}}
\label{table:1}
\end{table*}

\begin{table*}[ht!]
  \centering
  \renewcommand{\arraystretch}{3}
  \resizebox{5in}{!}{%
    \begin{tabular}{|c|c|c|}
      
      \hline
      \multicolumn{1}{|c|}{\textbf{\LARGE Data}} & \multicolumn{1}{c|}{\textbf{\LARGE Physical Parameters}} & \textbf{\LARGE Data Features} \\
      
      \hline
      \multicolumn{1}{|c|}{\multirow{4}{*}{\textbf{\Huge $D_2^\dagger$}}} & \textbf{\LARGE Chirp Mass ($M_{\odot}$)} & \LARGE $1.0$, $1.2$, $1.4$, $1.6$, $1.8$ \\
      \cline{2-3}
      \multicolumn{1}{|c|}{} & \textbf{\LARGE Mass Ratio} & \LARGE $0.7$, $0.75$, $0.8$, $0.85$, $0.9$ \\
      \cline{2-3}
      \multicolumn{1}{|c|}{} & \textbf{\LARGE Fraction of the Remnant Disk} & \LARGE $0.15$, $0.20$, $0.25$, $0.30$, $0.35$, $0.40$ \\
      \cline{2-3}
      \multicolumn{1}{|c|}{} & \textbf{\LARGE Viewing Angle} & \LARGE $45\degree$ ,$60\degree$, $75\degree$, $90\degree$ \\
      \hline
    \end{tabular}%
  }
\caption{This table contains the interpolated physical parameters that were used to simulate light curves using \textsc{MOSFIT} and compare those with the CVAE-generated light curves. Light curves from these physical parameters, alongside from the test set, are employed to evaluate the performance of the CVAE over the parameter space.}
\label{table:2}
\end{table*}

\section{Results}\label{sec:result}
In this section, we present the results after the implementation of CVAE. After training, samples are drawn from the latent space to reconstruct light curves based on the required physical parameters. This section points out the comparison between the generated light curves from the latent space and the original light  curves from the simulation. This technique allows us to generate as many light curves as we require over a wide range of physical parameters within the range provided by the model.

\begin{figure*}[htp!]
\gridline{\fig{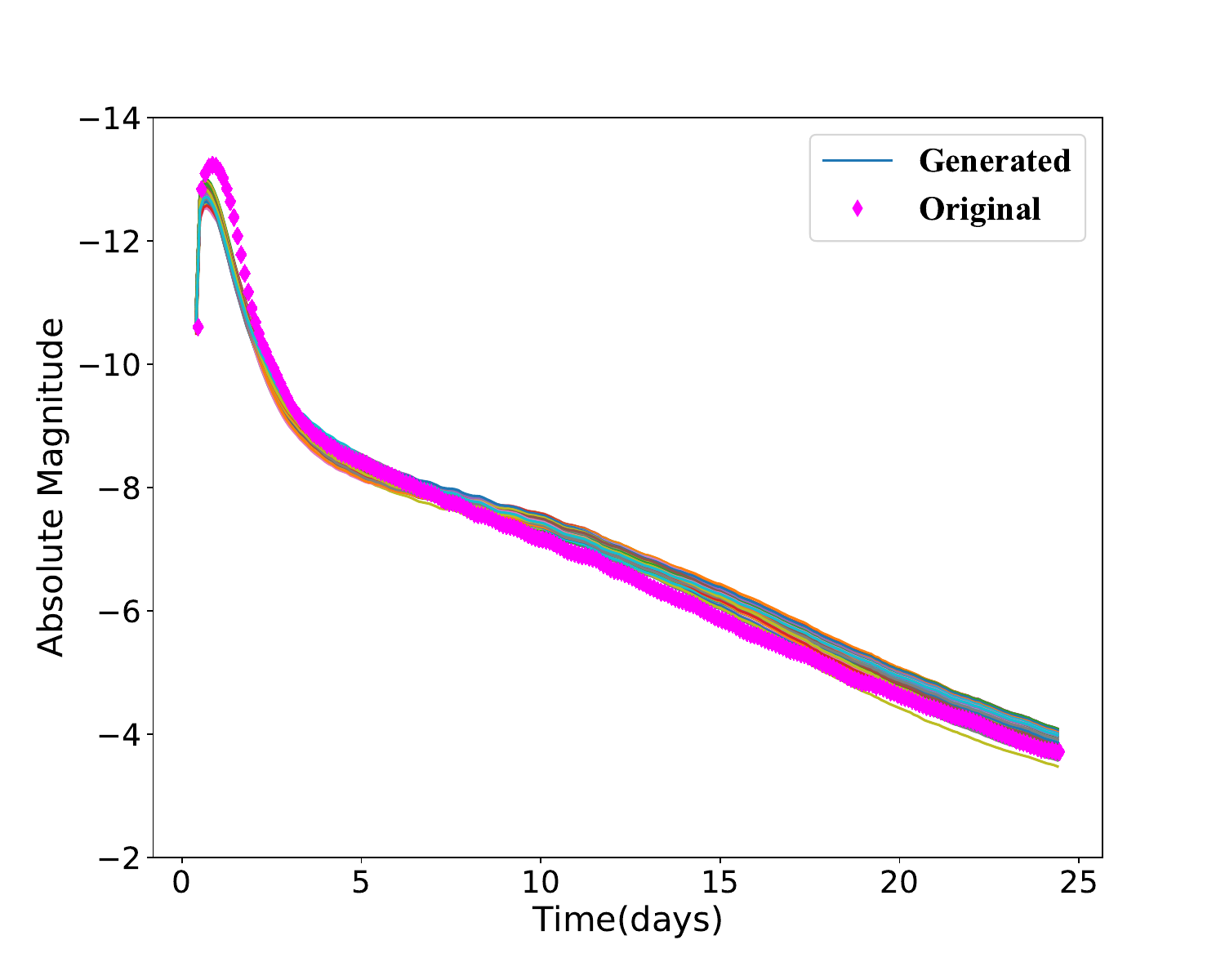}{0.5\textwidth}{(a)\hspace{0.1cm}g-band}
          \fig{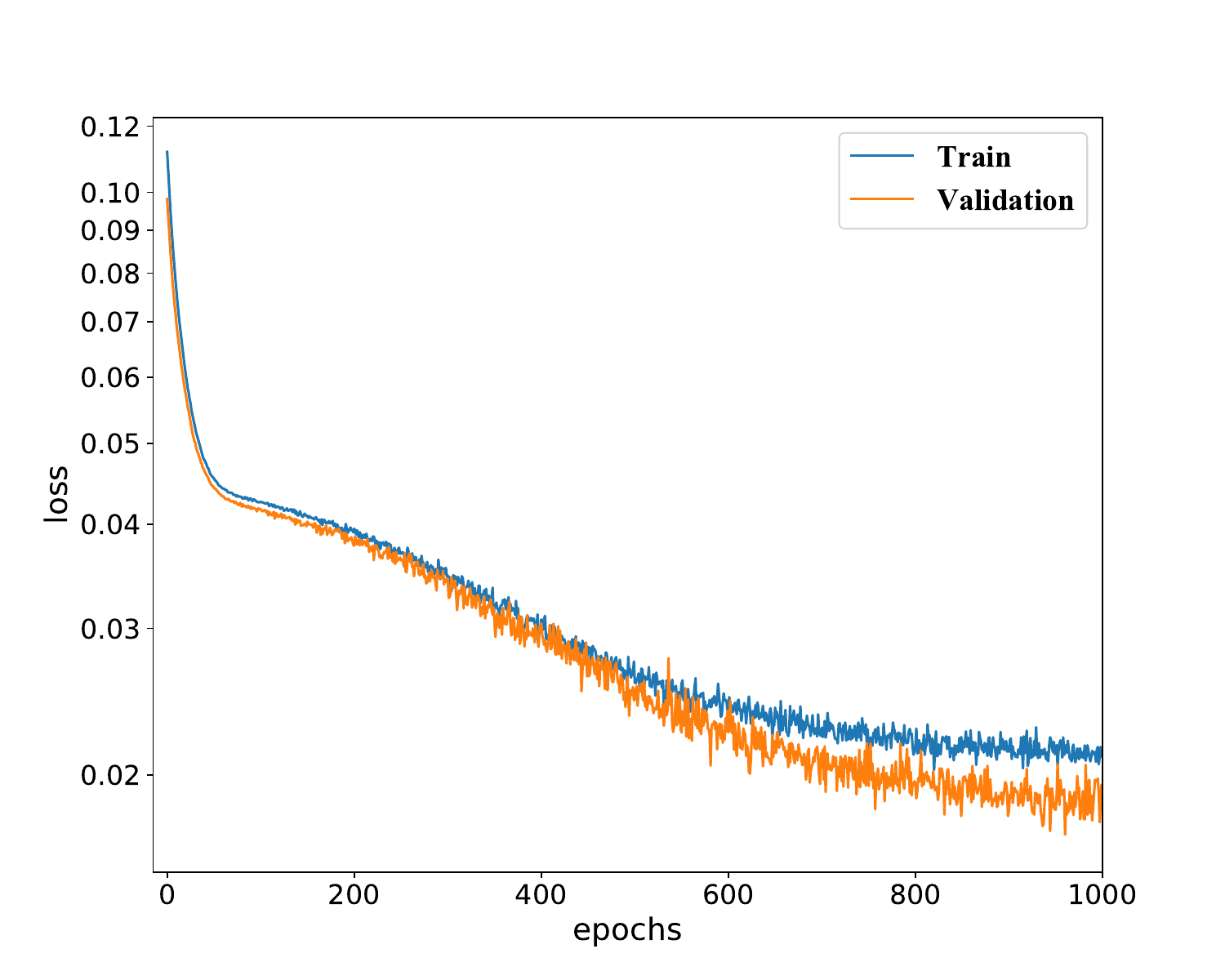}{0.5\textwidth}{(b)\hspace{0.1cm}Loss Plot}
          }\caption{(a)This figure show light curves comparison corresponding to g-band after training and generation are implemented on $D_{1}$.  We display $100$ light curves for physical parameter of ejecta mass, ejecta velocity and lanthanide fraction having values $[0.001M_{\odot},0.03c,10^{-5}]$ and compare it along side the light curve from the test data
          for the same physical parameter. The original light curve is shown in diamond marker and is within distribution of the generated light curves shown by dashed lines. The deviations seen in the generated light curves arise due to the variations in the latent space. (b)This plot corresponds to the learning curve after implementing the CVAE. The gap between the two curves is small indicating a good-fit}

\label{fig:cvaeresults1}          
\end{figure*}

In Fig.\ref{fig:cvaeresults1}(a), we show the generated light curves of g-band after implementing CVAE for the physical parameter of $[0.001M_{\odot},0.03c,10^{-5}]$ which are the ejecta mass, ejecta velocity, and lanthanide fraction, respectively as mentioned in Section~\ref{sec:data}. The loss plot for the CVAE is shown in Fig~\ref{fig:cvaeresults1}(b). We find that the validation loss and the training loss has decreased to a point of stability and there is small gap between the two curves. Although, CVAE has the ability to generate as many light curves we want for one or many physical parameters, but in order to avoid congestion in plots, we limit ourselves to $100$ generated light curves based on the above physical parameter. The $100$ light curves are closely spaced and the apparent variations at the later days comes from the variations of the probability distribution in the latent space. These variations are expected since, to regenerate light curves, we randomly draw samples from the latent space. This randomness allows the CVAE to produce variations in the light curves even for the same input. Thus, we see these variations in the CVAE-generated light curves for the physical parameters. Besides, the light curves are more sensitive to the changes in the physical parameters in the later days. However, since we are more interested in the earlier KNe evolution, we restrict the plots to $14$ days.
This parameter is chosen from the simulated data in order to verify the robustness of the CVAE. It is clear that the input light curve, shown by diamond markers, used from the validation data of the CVAE is well within the limit of the regenerated light curves. 

This assures that the CVAE is well trained. In Fig.~\ref{fig:cvaeresults1}(a), we see some overlapping in the light curves, since the $100$ generated light curves have relatively similar values. While generating these light curves, samples were drawn from different points in latent space.
It is possible to generate as many light curves as desired and observe the possible variations from the latent representation. Since the encoder maps the input light curves to a probability distribution over the latent space. Hence, even though in the respective KNe models, there is one light curve associated with a single value of the physical parameter, implementing CVAE provides a comparatively wide range of light curve distribution for the same value of physical parameter arising from the latent representation. This variation of the generated KNe light curves are particularly interesting when the light curves are generated for an entirely new physical parameter set which the CVAE has not come across during training and validation.

\begin{figure*}[htp!]
\plotone{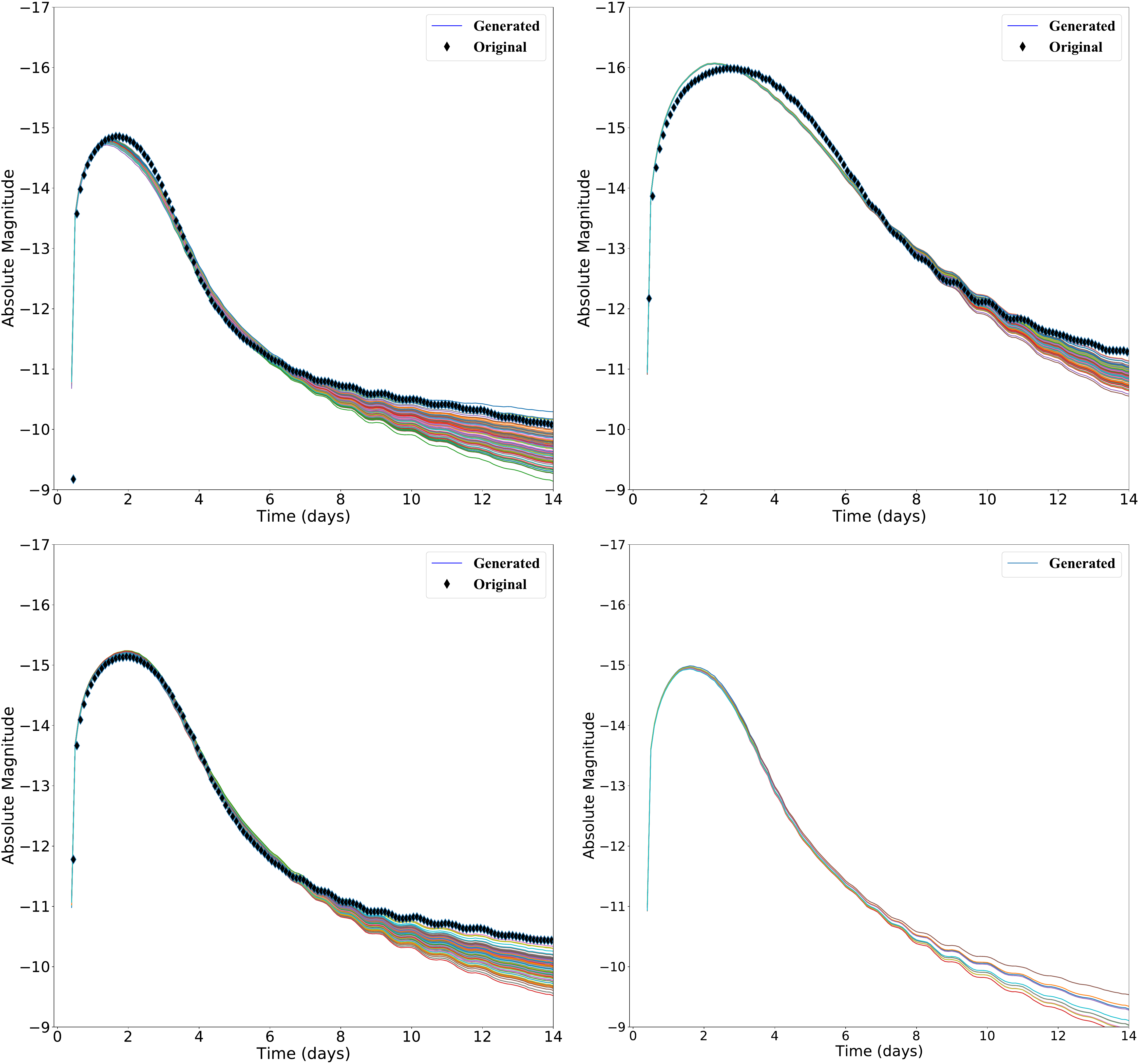}
\caption{In this figure, we compare $100$ light curves (solid lines) generated from CVAE with original the light curves (diamond marker) for physical parameter values having ejecta masses of $0.02M_{\odot}$(\emph{upper left}),$0.1M_{\odot}$(\emph{upper right}) and $0.03M_{\odot}$(\emph{bottom left}) keeping the ejecta velocity $0.03c$ and lanthanide fraction$(10^{-9})$ constant for all the generated light curves. These values of the physical parameter were chosen from the test set. In the bottom right panel, $10$ light curves for an arbitrary physical parameter $[0.025M_{\odot}$, $0.035c$, $10^{-5}]$ is shown. All the generated and the original light curves corresponds to the g-band.}

\label{fig:cvaeres2}
\end{figure*}

\begin{figure*}[hbt!]
\plotone{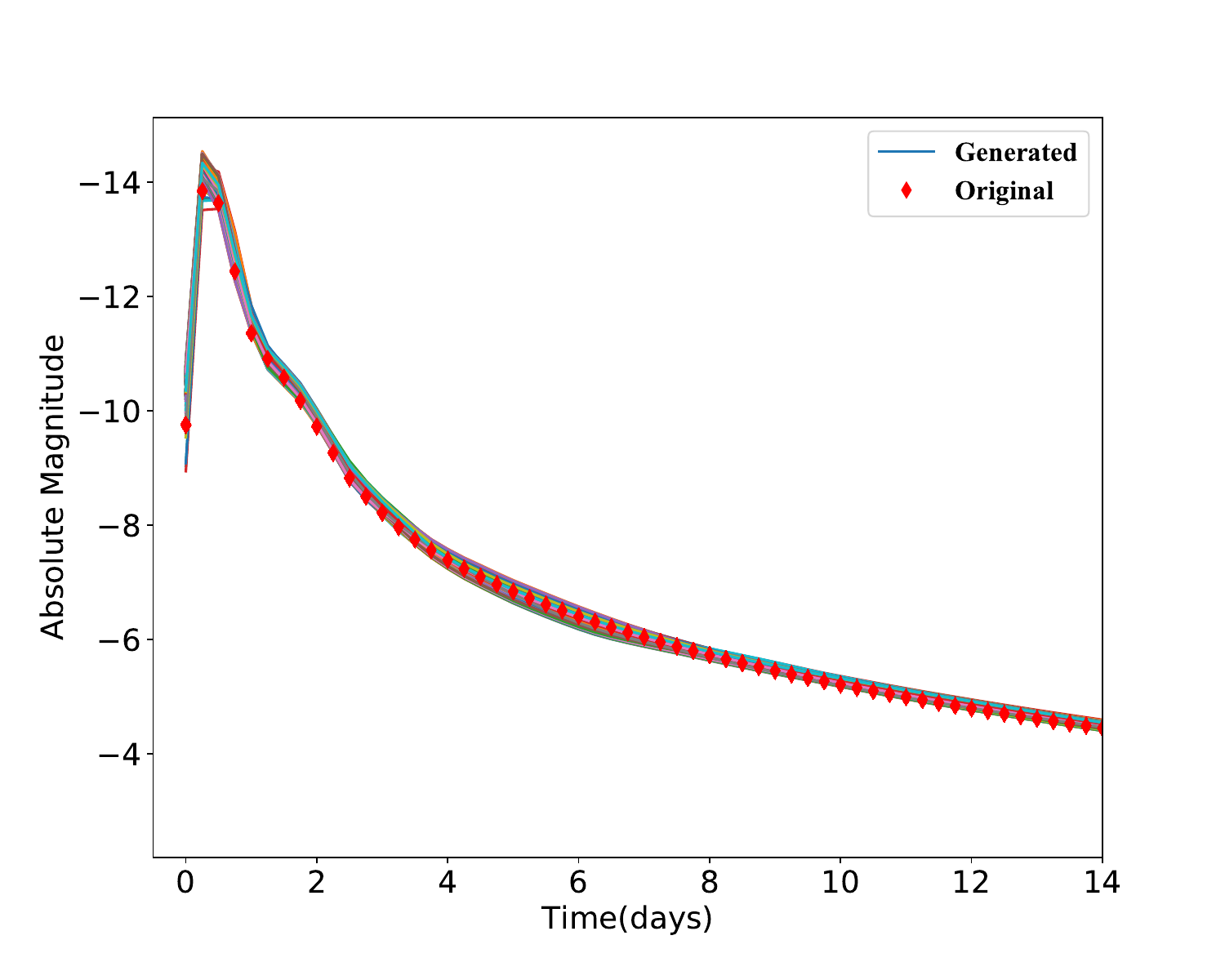}
\epsscale{1.2}
          \caption{Comparison for the true and generated light curves corresponding to $D_{2}$. These $100$ light curves are particularly from r-band generated from the CVAE where the comparison is shown with the respective physical parameter of chirp mass, mass ratio, fraction of the remnant disk and viewing angle from the pole having values of $[1.8M_{\odot},0.9,0.15,90\degree]$ respectively. The diamond marker corresponds to the true light curve, whereas the solid lines corresponds to the CVAE-generated light curves. The physical parameter for the true light curve is taken from the test set for verifying with the generated ones.}     
\label{fig:compM}
\end{figure*}
Kilonova light curves are extremely sensitive to ejecta mass\citep{2017Natur.551...80K}. With the change in the ejecta mass, the peak value of the light curves increases and shifts to later days, which is explicitly shown in the top left, top right and bottom left panel in Fig.\ref{fig:cvaeres2}. This result is expected in accordance to the KNe evolution. In this figure, the comparative analysis for the different values of ejecta masses $(0.02M_{\odot},0.1M_{\odot}$ and $0.03M_{\odot})$ are shown keeping the ejecta velocity $(0.03c)$ and lanthanide fraction $(10^{-9})$ unchanged. $100$ light curves are generated from the CVAE comparing with the original light curve for the above physical parameters. In the bottom right panel, light curve for an arbitrary parameter $(0.025M_{\odot}$,$0.035c$,$10^{-5})$ is shown. This physical parameter value configuration in not present in $D_{1}$ therefore we do not have any benchmark to validate it, however, with more data from the respective KNe model, this can be verified. Similar comparative analysis, for all other filter bands, showing the evolution of the light curves with the change in ejecta mass, ejecta velocities and lanthanide fractions are shown in appendix in Fig~\ref{fig:app1}. 
We next take the second data set, $D_{2}$, with physical parameters of chirp mass, mass ratio, fraction of the remnant disk and viewing angle from the pole. We run the same analysis as above. As mentioned in Section~\ref{sec:data}, this data set has a total of $529$ light curves, with the above physical parameters having values given as mentioned in Section~\ref{sec:data}. The training data has $401$ light curves, and the remaining are equally divided into a test set and a validation set. In Fig~\ref{fig:compM}, we demonstrate the application of the CVAE on $D_{2}$ where we have compared $100$ light curves generated from CVAE with the original light curve for the physical parameter of $[1.8M_{\odot},0.9,0.15,90\degree]$ correlated to the chirp mass, the mass ratio, the fraction of the remnant disk and the viewing angle from the pole, respectively. This set of physical parameters is taken from the test data set

\begin{figure*}[h!]
\centering
\epsscale{1.2}
\plotone{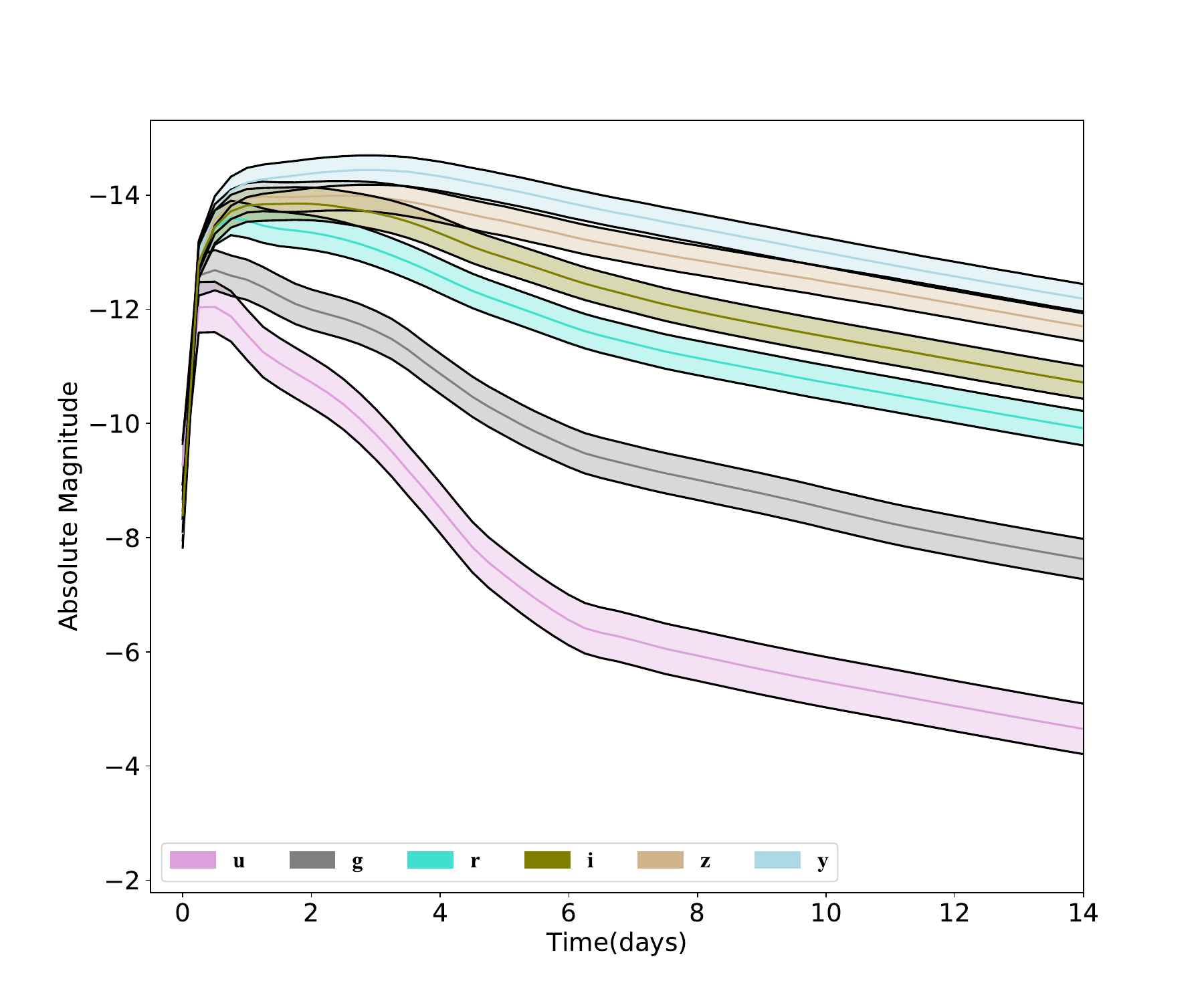}

\caption{This figure corresponds to the $90\%$ confidence plot for the CVAE-generated and true light curves in the \textit{g,r,z,y,i} and \textit{u} bands that have the physical parameter of $1.2M_{\odot}$ chirp mass, $0.7$ mass ratio, $0.15$ fraction of the remnant disk and a viewing angle of $60\degree$. True values of the light curves for the above parameter are represented by the solid lines in each color-filled region. We find quite satisfactory agreement between the true and CVAE-generated light curves. 
The light curves corresponding to the above parameters were taken from the $D_{2}\dagger$ data set and thus are entirely new to the CVAE for prediction and generation.}
\label{fig:view_comp}
\end{figure*}

The input light curve is within the distribution of the generated light curve which indicates that the CVAE is performing well and the training and regeneration of the light curves are successful.
Fig.~\ref{fig:view_comp}, shows the confidence plot for a set of physical parameter where it is important to mention that the neural network has not seen this set of physical parameter and hence from the similarity between the generated and true light curves we can substantiate the performance of the CVAE. Other relevant plots for different combination of physical parameter and the changes in the light curves corresponding to those for other filter bands are shown in Fig~\ref{fig:app2} in appendix. To measure the predictive accuracy of the trained CVAE, we use mean absolute error (MAE), where lower value corresponds to more accurate prediction, calculated using
\begin{equation}\label{eq:1}
    MAE=\frac{1}{n}\sum_{i=1}^{n}|y_i-\hat{y}_i|
\end{equation}, 
where $y_i$ and $\hat{y}_i$ are the true values and predicted values of the light curves of the test data and $n$ is the total number of points drawn for each light curve from latent space. Alongside the above mean absolute error, we also calculate and plot the mean squared error (MSE) between the true and predicted  light curve values using the equation
\begin{equation}\label{eq:2}
    MSE=\frac{1}{n}\sum_{i=1}^{n}(z_i-\hat{z}_i)^2
\end{equation}
where, $n$ being the total number of data points and $z_i$ and $\hat{z}_i$ are the respective values of true and CVAE-predicted light curves.

\begin{figure*}[ht!]
\centering
\epsscale{1.2}
\plotone{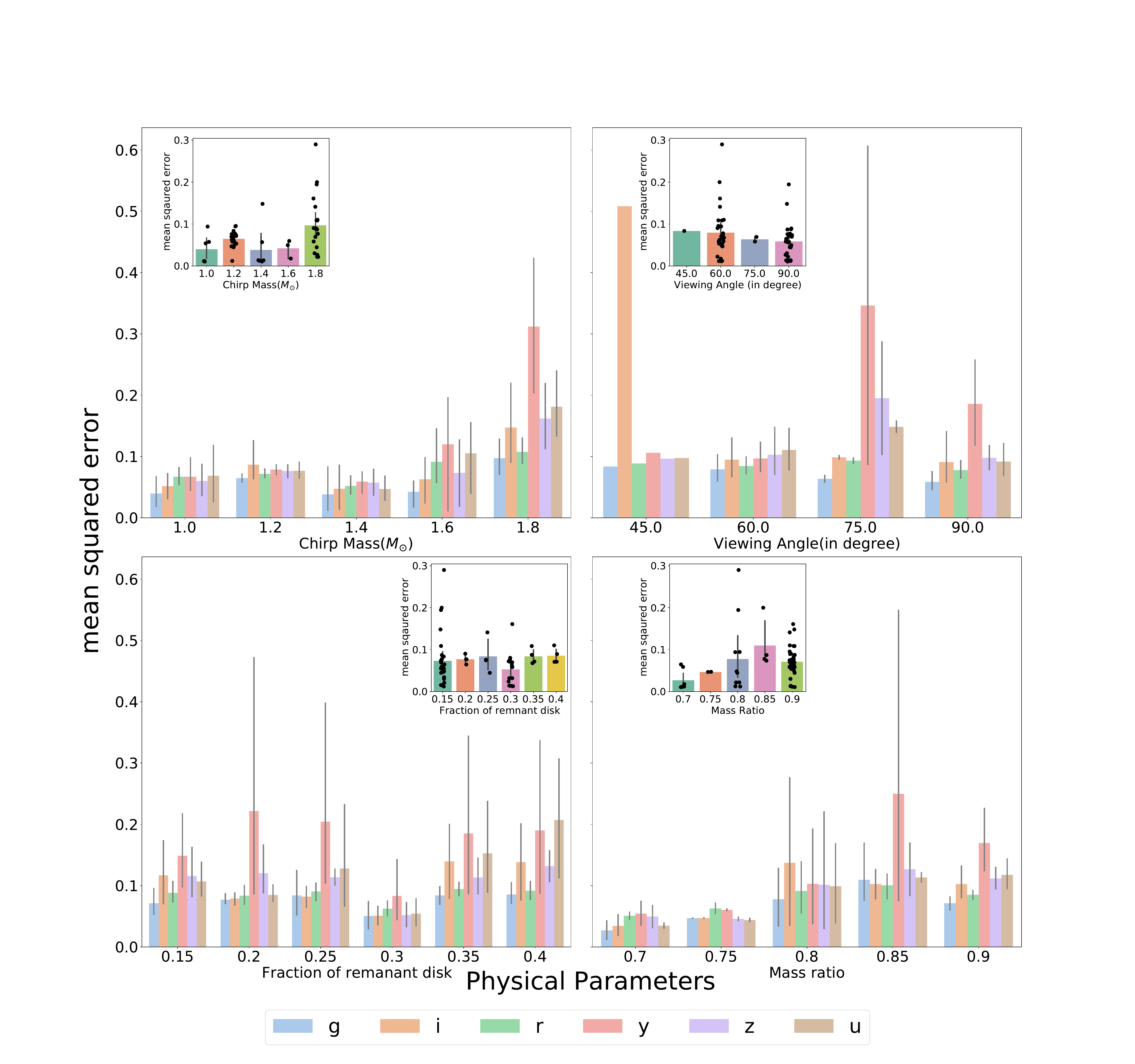}
\caption{In this figure, we present the performance of the CVAE spanning over the whole parameter space of chirp mass $(1.0M_{\odot}-1.80M_{\odot})$,viewing angle$(45\degree-90\degree)$,fraction of remnant disk$(0.15-0.40)$ and  mass ratio $(0.70-0.90)$ in the \emph{u,g,i,r,z} and \emph{y} filter bands represented by the different colors bars grouped accordingly. The \emph{y-axis} represents the mean squared values whereas in the \emph{x-axis} we have the respective physical parameter values for which we have calculated the overall mean squared error in the different filter bands shown with different color bars.  For the chirp mass frame(\emph{upper left}),  the mean squared errors for different filer bands are grouped by the chirp mass values, $1.0M_{\odot},1.2M_{\odot},1.4M_{\odot},1.6M_{\odot},1.8M_{\odot}$, while covering parameter space of viewing angle, mass ratio and fraction of the remnant disk. For the viewing angle frame(\emph{upper right}), we show the error for $45\degree$, $60\degree$, $75\degree$ and $90\degree$ viewing angles calculated over the entire parameter values of chirp mass, mass ratio and fraction of the remnant disk. In the remnant disk frame(\emph{lower left}), for the values of $0.15$,$0.20$,$0.25$,$0.30$,$0.35$ and $0.40$ fraction of remnant disk, errors are grouped in different filter bands encompassing the chirp mass, viewing angle and mass ratio. For the mass ratio frame(\emph{lower right}), errors corresponding to mass ratio values of $0.70$,$0.75$,$0.80$,$0.85$,and $0.90$ are shown as grouped bar plots across the parameter space of chirp mass, viewing angle and fraction of the remnant disk for the different filter bands.  In the insets for all the frames, we show the CVAE performance over the entire parameter space, as discussed above, obtained from CVAE generated \emph{g-band} light curves. Each dot in the inset corresponds to the calculated value of the mean squared error of a light curve for the relevant set of physical parameters. For each of the histograms corresponding to the different parameter space, error bars have been shown. The histograms without the error bar correspond to a single light curve that is available for calculating the mean squared error between the true and CVAE generated light curves.}
\label{fig:cvae_perf}
\end{figure*} 
\section{Discussion} \label{sec:dis}
Since the CVAE depends on the input data, light curves from different KNe models can be used for training and the corresponding trained CVAE is saved for generating light curves.The input data contains both training data and physical parameters. This versatility of the technique provide a unique opportunity to test this on other data set. Using eq(\ref{eq:1}), we have the mean absolute error to be $0.0995$ and $0.0339$ for $D_{1}$ and $D_{2}$ respectively. The mean squared error for $D_{1}$ and $D_{2}$ calculated on the test data using eq(\ref{eq:2}) are found to be $0.00241$ and $0.00153$ respectively.

Compared to $D_{2}$, the MAE for $D_{1}$ is higher, as the training data is relatively lower for $D_{1}$.
We not only successfully reconstructed the light curves for known values of physical parameters but also generated light curves for the physical parameters that are not present in the data set but are well within the range of the physical parameters values provided in the respective models.
In Fig.~\ref{fig:cvaeresults1}, we presented the reconstructed light curves and the corresponding loss values for the CVAE. Consequently, based on these results, we proceed to generate light curves for different combinations of physical parameters and verify those as presented in Fig.~\ref{fig:cvaeres2}. The obtained results are consistent with the KNe evolution presented in the model. Comparing the plots in Fig.~\ref{fig:cvaeresults1}(a) and Fig.~\ref{fig:cvaeres2}, as mentioned in Section~\ref{sec:result}, we can see more variations in the latter, and for the latter figure the light curves has been truncated to $14$ days while in the former, CVAE-generated light curves are extended till the end.      
For the plots generated from $D_{2}$, since there has been no significant change related to astrophysical importance, we have truncated the light curves to $14$ days. Model evaluation is performed with the test data, in the current work, we move a step ahead to evaluate the performance via the confidence interval (Fig.~\ref{fig:view_comp}) and mean squared error (Fig.\ref{fig:cvae_perf}) between the CVAE-generated and true light curves by taking physical parameters from $D_2\dagger$ which contains the physical parameters from the test set as well as the entirely new physical parameters that were absent in the training, test and validation set. Therefore, as previously mentioned in Section.~\ref{sec:data}, we use the physical parameters combination from Table~\ref{table:2} to produce Fig.~\ref{fig:cvae_perf}. Therefore, when evaluating the performance on these sets of previously unseen parameters, simulates the real scenario akin to observing a new kilonova.
In Fig.~\ref{fig:view_comp}, we represent the figure for the $90\%$ confidence interval between the generated and true light curves in different filter bands of \textit{u,g,r,i,z} and \textit{y} having the chirp mass of $1.2M_{\odot}$, mass ratio of $0.7$, $0.15$ fraction of the remnant disk having a viewing angle of $60\degree$. The true light curves for this set of parameters were not included in the training, test, or validation data set, thus this data is unseen to the network. To plot the confidence interval, we take the mean light curves from the $2000$ CVAE-generated light curve for the above physical parameter values [$1.2M_{\odot}$,$0.7$,$0.15$,$60\degree$] in each filter bands. Each result is highlighted with color-shaded region for each filter band including the true light curve. Here we see a good agreement between the true and generated light curves which provides supporting evidence in favour of the performance of CVAE. In addition to the above, we have calculated the mean squared error between the generated and true light curves where we have have obtained the error to be always between $(0.015-0.08)$ for the set of physical parameters. To provide further evidence of CVAE performance, we illustrate the mean squared error calculated  over the entire parameter space while considering the following values of physical parameters,$1.0M_{\odot}$,$1.2M_{\odot}$,$1.4M_{\odot}$,$1.6M_{\odot}$, and $1.8M_{\odot}$ for the chirp mass region, $45\degree$, $60\degree$, $75\degree$ and $90\degree$ for viewing angle region, $0.15$, $0.20$, $0.25$, $0.30$, $0.35$ and $0.40$ for the fraction of remnant disk parameter and for the $0.70$, $0.75$, $0.80$, $0.85$ and $0.90$ in mass ratio parameter for the different filter bands as shown in Fig.\ref{fig:cvae_perf}. To calculate the mean squared error, we have used the mean of $2000$ CVAE-generated light curves and the true light curves for the respective sets of physical parameters. \par 
Fig.\ref{fig:cvae_perf} consists of four frames, each having an inset, which show the error calculated for the respective value of physical parameter (along \emph{x-axis}) across the parameter space of the other remaining physical parameters. For the chirp mass frame (\emph{upper left}),error is calculated for the different chirp masses ($1.0M_{\odot}$,$1.2M_{\odot}$,$1.4M_{\odot}$,$1.6M_{\odot}$,$1.8M_{\odot}$) across the parameter space of viewing angle, mass ratio and the fraction of the remnant disk, grouped together for each value of chirp mass for the filter bands. For the viewing angle frame (\emph{upper right}), for the values of $45\degree$,$60\degree$,$75\degree$ and $90\degree$ error is calculated across the chirp mass, mass ratio and fraction of the remnant disk and grouped accordingly. For calculating the mean squared error for the fraction of remnant disk, we have considered $0.15$,$0.20$,$0.25$,$0.30$,$0.35$,and $0.40$ across the parameter space of chirp mass, viewing angle and mass ratio grouped in parameter values in different filter bands. In the mass ratio frame (\emph{lower right}), error is calculated for the mass ratio values of $0.70$,$0.75$,$0.80$,$0.85$, and $0.90$, across the parameter space of chirp mass, viewing angle and fraction of the remnant disk grouped accordingly. For certain set of physical parameter in certain filter bands, tend to have comparatively high mean squared error value, thus adding more to the overall error. For instance, the above case can be seen in the viewing angle frame in Fig.\ref{fig:cvae_perf} for $45\degree$ viewing angle in \emph{i-band}. However, the overall error are considerably low. In the insets of Fig.\ref{fig:cvae_perf}, \emph{g-band} results are shown where each dot refers to the mean squared error of a single light curve for the respective set of physical parameter across the parameter space. Here also, we find that for certain set of physical parameter, the error value is comparatively higher as evident from outliers in $1.8M_{\odot}$ chirp mass, $60\degree$ viewing angle, $0.15$ fraction of remnant disk and $0.80$ mass ratio in the insets. In comparison to one-one mean squared error, the overall error is relatively higher as it includes all the parameter sets across the individual parameter considered. From the above result, we find the regions in parameter space for different filter bands where the CVAE has comparatively unsatisfactory results. In addition to the above, the network summary, Fig~\ref{fig:app3} has been added in the appendix for reference. While comparing Fig.~\ref{fig:cvaeresults1} and Fig.~\ref{fig:cvaeres2} for $D_1$ with results from $D_2$ in Fig~.\ref{fig:compM}, one will find similar deviations in the tail of the light curves after $10$ days for the different sets of physical parameters shown in plots presented in the appendix(Fig~\ref{fig:app2}). In the main text, only one such plot (Fig.~\ref{fig:compM}) has been shown for a single set of physical parameter, where the deviations are comparatively less. One of the main reasons for such different deviations is distribution of the latent space owing to the difference in the number of training data, since $D_1$ has comparatively less training data than $D_2$.

\section{Conclusion}\label{sec: con}
In this paper, we look into rapid generation of KNe light curves based on different physical parameters values while implementing CVAE. We present a methodological approach to our idea. In the initial stages, the performance check was carried out while inspecting the loss curves during the training and validation. We also evaluated the CVAE performance on different physical parameter from test data, hence perform rapid interpolation of light curves and subsequently provide evidence that the results are quite in agreement with the true light curves. However, in certain cases, while generating the light curves, we find that for $D_1$, as compared to the original light curves, the generated light curves tends to deviate after $8$ days but for $D_2$ we do not see such deviations in the later days. Here, we have used publicly available KNe data to provide a proof of our concept, but this does not put any limitation on this technique. This technique can also be used for similar kind of data analysis comprising of data related to other domain in astronomy and astrophysics. The striking point for such an approach is that it reduces the time which is required by simulations to rerun and reproduce similar results for different physical parameters every time. 
We train the data available and produce the desired results rapidly rather than adjusting the simulation code each time and the saved CVAE model can be used to generate the required KNe light curves.In this work, by speeding up the light curve generation by $1000$ times, we have achieved rapid results. Besides the current application for KNe, the CVAE technique demonstrated above has prospective applications in a similar procedure where data are available for training, test and, validation. In the current work, since the detailed calculations of the simulation are not incorporated into the CVAE, we do not expect any new results, but at the same time, the trained-CVAE produces results without looking into the particulars of the simulation.An alternative approach for generating simulation results, with the help of machine learning tools, without actually re-executing those is presented here.Additionally, this method can accommodate other KNe models where data are available to be fed into the network. This kind of CVAE approach has also the potential to be utilized for rapid parameter estimation not only limiting to KNe but also for other astrophysical sources where rapid data analysis is required.

\section{Acknowledgements} This work is supported by the National Science and Technology Council of Taiwan under the grants 110-2628-M-007-005 and 111-2112-M-007-020, and a joint grant of the National Science and Technology Council and the Royal Society of Edinburgh through 110-2927-I-007-513. MJW is supported by the Science and Technology Facilities Council [2285031,ST/V005634/1,ST/V005715/1]. ISH is supported by the Science and Technology Research Council [ST/ L000946/1]. ISH and MJW are also supported by the European Cooperation in Science and Technology (COST) action [CA17137]. MN is supported by the European Research Council (ERC) under the European Union’s Horizon 2020 research and innovation programme (grant agreement No.~948381) and by funding from the UK Space Agency. The authors are grateful to the valuable suggestions from He-Feng Hsieh, John Veitch, and Nicola De Lillo. 

\software{matplotlib \citep{Hunter:2007}, pandas\citep{reback2020pandas}, tensorflow\citep{tensorflow2015-whitepaper}, keras\citep{chollet2015keras} MOSFIT \url{https://github.com/mnicholl/MOSFiT, https://github.com/guillochon/MOSFiT}}
\clearpage
\nocite{*}
\bibliography{sample631}{}
\bibliographystyle{aasjournal}

\appendix
\label{app}
\restartappendixnumbering
\section{Additional Results from $D_{1}$}
\begin{figure*}[ht!]
\centering
\epsscale{1.15}
\plotone{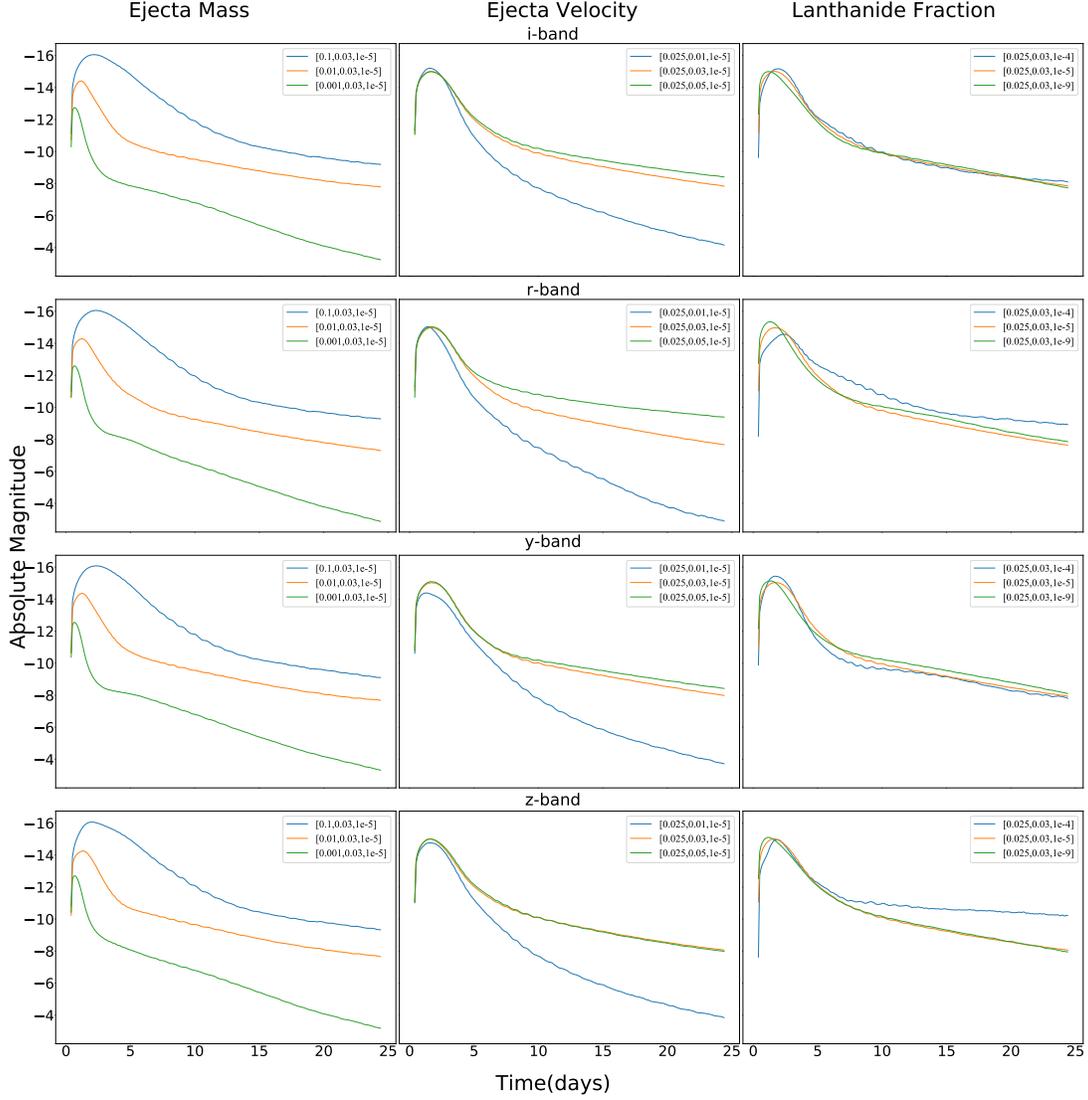}
\caption{This figure encompasses the CVAE-generated light curves for different values of physical parameters for the respective filter bands. In each row of the figure, light curves corresponds to different ejecta mass having values of $(0.1M_{\odot},0.01M_{\odot},0.001M_{\odot})$keeping ejecta velocity $(0.03c)$ and lanthanide fraction $(10^{-5})$ constant; different ejecta velocity having values $(0.01c,0.03c,0.05c)$ keeping ejecta mass$(0.025M_{\odot})$ and lanthanide fraction $(10^{-5})$ to be constant; different lanthanide fraction having values $(10^{-4},10^{-5},10^{-9})$, keeping ejecta mass $(0.025M_{\odot})$ and ejecta velocity $(0.03c)$ to be constant for \emph{i,r,y} and \emph{z} filter bands respectively.}
\label{fig:app1}
\end{figure*}
\clearpage
\restartappendixnumbering
\section{Additional Results from $D_{2}$}
\begin{figure*}[ht!]
\centering
\epsscale{1.19}
\plotone{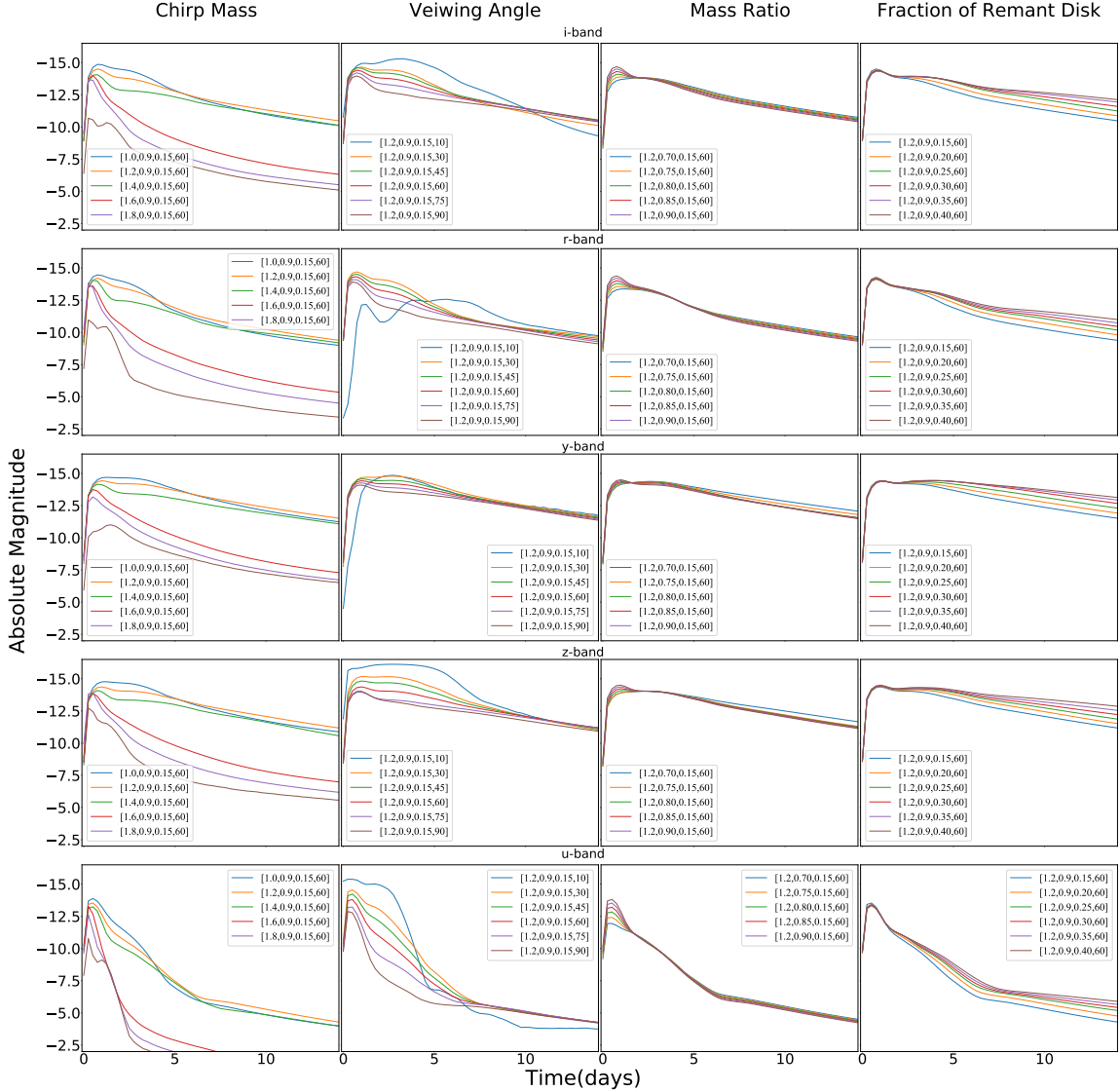}
\caption{This figure demonstrates the evolution of light curves for different physical parameters of chirp mass, viewing angle, mass ratio and fraction of the remnant disk. In each row of the plot, we bring together the CVAE-generated light curves for different chirp mass values $(1.0M_{\odot})$,$(1.2M_{\odot})$,$(1.4M_{\odot})$,$(1.6M_{\odot})$ and $(1.8M_{\odot})$ while keeping the viewing angle($60\degree$), mass ratio($0.9$) and fraction of the remnant disk($0.15$) to be constant; for different viewing angles having values $(10\degree, 30\degree, 45\degree, 60\degree, 75\degree, 90\degree)$ from the pole keeping the chirp mass $(1.2M_{\odot})$, mass ratio$(0.9)$ and the fraction of the remnant disk ($0.15$) constant; for different fraction of the remnant disk having values $(0.15,0.20,0.25,0.30,0.35,0.40)$, keeping the chirp mass$(1.2M_{\odot})$, mass ratio $(0.9)$ and the viewing angle $(90\degree)$ constant for the \emph{i,r,y,z} and \emph{u} filter bands respectively.}
\label{fig:app2}
\end{figure*}

\restartappendixnumbering
\section{The CVAE Summary.}
\begin{figure*}[ht!]
\centering
\epsscale{1}
\plotone{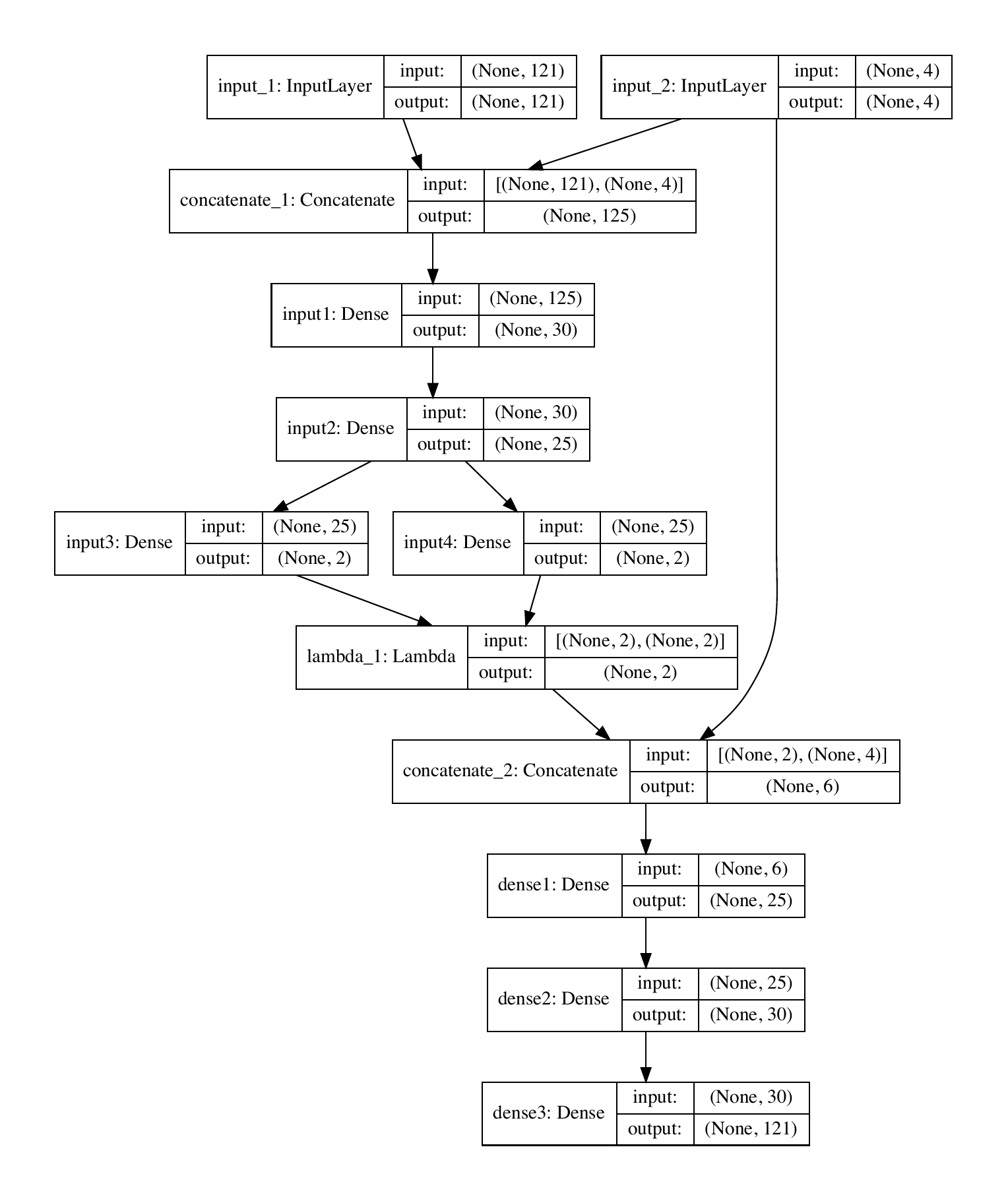}
\caption{This is the summary of the CVAE implemented in the current work}
\label{fig:app3}
\end{figure*}

\end{document}